\documentclass[preprint,showpacs,preprintnumbers,amsmath,amssymb,superscriptaddress]{revtex4}
\usepackage{graphicx}
\usepackage{dcolumn}
\usepackage{bm}

\begin{document}

\title{High-Q microcavity enhanced optical properties of CuInS$_{2}$/ZnS colloidal quantum dots towards non-photodegradation}

\author{Yue Sun}
\author{Feilong Song}
\author{Chenjiang Qian}
\author{Kai Peng}
\author{Sibai Sun}
\author{Yanhui Zhao}
\affiliation{Beijing National Laboratory for Condensed Matter Physics, Institute of Physics, Chinese Academy of Sciences, Beijing,
100190, China}
\author{Zelong Bai}
\affiliation{Beijing Key Laboratory of Nanophotonics and Ultrafine Optoelectronic Systems, School of Materials Science $\&$
Engineering, Beijing Institute of Technology, Beijing, 100081, China}
\author{Jing Tang}
\author{Shiyao Wu}
\author{Hassan Ali}

\affiliation{Beijing National Laboratory for Condensed Matter Physics, Institute of Physics, Chinese Academy of Sciences, Beijing,
100190, China}
\author{Fang Bo}

\affiliation{The MOE Key Laboratory of Weak Light Nonlinear Photonics, TEDA Applied Physics Institute and School of Physics, Nankai University, Tianjin 300457, China}

\author{Haizheng Zhong}

\affiliation{Beijing Key Laboratory of Nanophotonics and Ultrafine Optoelectronic Systems, School of Materials Science $\&$
Engineering, Beijing Institute of Technology, Beijing, 100081, China}
\author{Kuijuan Jin}
\affiliation {Beijing National Laboratory for Condensed Matter Physics, Institute of Physics, Chinese Academy of Sciences, Beijing, 100190, China}

\affiliation {School of Physical Sciences, University of Chinese Academy of Sciences, Beijing 100190, China}
\affiliation {Collaborative Innovation Center of Quantum Matter, Beijing, China}
\author{Xiulai Xu}
\email{xlxu@iphy.ac.cn}
\affiliation{Beijing National Laboratory for Condensed Matter Physics, Institute of Physics, Chinese Academy of
Sciences, Beijing, 100190, China}
\affiliation{School of Physical Sciences, University of Chinese Academy of Sciences, Beijing 100190, China}

\begin{abstract}
We report on a temporal evolution of photoluminescence (PL) spectroscopy of CuInS$_{2}$/ZnS colloidal quantum dots (QDs) by drop-casting on SiO$_{2}$/Si substrates and high quality factor microdisks (MDs) under different atmospheric conditions. Fast PL decay, peak blueshift and linewidth broadening due to photooxidation have been observed at low excitation power. With further increasing of the excitation power, the PL peak position shows a redshift and linewidth becomes narrow, which is ascribed to the enhanced F$\ddot{o}$rster resonant energy transfer between different QDs by photoinduced agglomeration. The oxygen plays an important role in optically induced PL decay which is verified by reduced photobleaching effect in vacuum. When the QDs drop-casted on MDs, photooxidation and photobleaching are accelerated because the excitation efficiency is greatly enhanced with coupling the pumping laser with the cavity modes. However, when the emitted photons couple with cavity modes, a PL enhancement by more than 20 times is achieved because of the increased extraction efficiency and Purcell effects of MDs at room temperature (RT), and 35 times at 20 K. The photobleaching can be avoided with a small excitation power but with a strong PL intensity by taking advantages of high quality factor cavities. The high efficient PL emission without photodegradation is very promising for using CuInS$_{2}$ QDs as high efficient photon emitters at RT, where the photodegradation has always been limiting the practical applications of colloidal quantum dots.
\end{abstract}

\maketitle
\section{Introduction}
Comparing to the traditional II-VI  colloidal quantum dots (QDs) \cite{Klimov1998,Klimov2000,Guo2003,Sark2001,Nazzal2004}, copper indium sulfide (CIS) QDs \cite{Chen2012,Li2013,Akdas2015,Sun2016}, also called nanocrystals, have been investigated intensively as an alternative for having less toxic materials such as cadmium or lead. Because of the large Stokes shift of about 300 meV \cite{Shabaev2015} and the wide emission energy tunability \cite{Chen2012}, CIS QDs have been explored in many applications such as light-emitting diodes \cite{Chen2012}, photovoltaics \cite{Pan2014}, and bioimaging \cite{Zhao2015}. CIS QDs based light emitting diodes have been demonstrated with a high electroluminescence intensity and a spectrum range from visible to near-infrared region \cite{Chen2012,Knowles2016}. Recently, a high photoluminescence (PL) quantum yield up to 86\% has been demonstrated in core/shell structured QDs \cite{Li2011}. Up to now, the radiative decay mechanism of CIS QDs is not very well understood \cite{Leach2016,Knowles2016,Whitham2016,Berends2016}, which could be the recombination from conduction band to localized intra-gap state \cite{Li2011,Berends2016,Knowles2015}, donor-acceptor pair (DAP) \cite{Zhong2012,Shi2012}, or from a localized state to valence band \cite{Omata2014}. In addition, instability of fluorescence always limits the colloidal QDs for commercialization because of the photooxidation and photobleaching, which has been studied mainly in conventional II-VI QDs \cite{Sark2001,Nazzal2004,Manner2012}. However, photodegradation of CIS QDs has rarely been investigated so far \cite{Whitham2016,Zhang2015}. Therefore, understanding the photoinduced degradation mechanism of CIS QDs systematically is highly desirable for a practical utilization of colloidal QDs.

Optical microcavities have been used to implement light-matter interaction because of the strong light confinement \cite{Vahala2003,Tang2015}, such as investigating cavity quantum electrodynamics and obtaining the emitted photons in desired directions. When the colloidal QDs are coupling with microcavities, enhanced photon emission from colloidal QDs have been observed with optical cavities \cite{Zhang2008,Yang2008,Kahl2007,Gupta2013,Abraham2009,Martiradonna2008,Foell2012,Ganesh2007,Rodarte2012} and plasmonic cavities \cite{Rakovich2015,Hoang2016} by greatly suppressing the spontaneous emission lifetime. Strong coupling between an exciton in a QD and a photon in a high quality factor (Q) optical resonator has been reported recently \cite{Yoshie2004,Reithmaier2004,Hennessy2007,Srinivasan2007,Brossard2010} when the exciton energy is resonant with the cavity mode. Among these microcavities, microdisks (MD) are relatively easy to be fabricated and have a very high Q of whispering-gallery modes (WGMs) \cite{Srinivasan2007,Bo2014,Xie2016}. Due to the large size of the MDs, they provide a great opportunity to drop-cast QDs on MDs for investigating to enhance optical properties of colloidal QDs.

To couple the colloidal QDs with the MDs, QDs have been embedded inside the MD cavity to maximize the coupling strength due to the better field overlapping \cite{Kahl2007,Xie2016}. However, this brings the fabrication processes of MDs much more complex and also limits the Q because of the non-uniformity of QD layer. In fact, the QDs could couple efficiently the evanescent electric field of cavity modes when QDs are drop-casted on top of the cavities, which also enhances the optical properties of QDs \cite{Ganesh2007,Foell2012}. In this work, we report on temporal evolutions of the PL properties of colloidal CIS/ZnS QDs dispersed on SiO$_{2}$/Si substrates and MDs. The PL intensity of QDs on the edge of MDs is greatly enhanced by 20 times at room temperature (RT) and 35 times at 20 K because of the Purcell effect of MDs. It is found that the PL intensity of close-packed CIS/ZnS QDs initially decreases and then keeps stable with a CW laser illuminating, which can be ascribed to photobleaching and photooxidation. The photoinduced degradation is investigated under air and vacuum atmospheric environments at different temperatures. We found the optically induced degradation can be avoided with a small excitation power by taking advantages of enhanced PL emission using MDs.

\section{Results and discussion}

Figure 1 shows schematic diagram of a conventional confocal micro-PL system for PL measurement of QDs on SiO$_{2}$/Si substrates and MDs (as shown in the inset). Figures 2(a)-2(c) show wide-field PL images of CIS QDs on a MD by shining a pumping laser at different locations. During the PL mapping, the pumping laser was filtered out with a long-pass filter. When the pumping laser touches the edge of the MD, as shown in Figure 2 (a), the contacting part turns red due to the PL, while the opposite MD edge away from the laser spot remains dark. When the laser totally irradiates on the edge of MD, the whole edge of the MD turns on, which is demonstrated in Figure 2 (b). This indicates that pumping laser couples to the WGMs of the MD, which excites the QDs on the edge. Another possibility is that the emitted light from QDs also couples to the cavity modes. After the laser goes far away from the edge of the MD, the PL intensity becomes weak as shown in Figure 2(c). To illustrate the cavity modes of the MD, a two dimensional finite-difference time-domain method was used to simulate the cavity modes. Figure 2 (d) shows electric field magnitude distribution of a TE mode of about 669.75 nm, which is close to the center of PL spectrum. The enlarged cavity mode distribution is shown in the inset of Figure 2(d). It can be seen the simulated field distribution on the edge of MD corresponds well with the observed PL images.

To get detailed spectral information, the PL spectra were mapped by moving the objective in a step of 1.5 $\mu$m. Figures 3(a) and 3(c) show the PL mapping of the QDs of a quarter of an area of the MD at RT and 20 K. The red annulus similar to the edge of the MD shows an enhancement of PL, which is induced by the cavity modes of the MD. The PL spectra from the edge of MD and from outside of MD are shown in Figures 3(b) and 3(d). The integrated PL intensity on the edge of the MD (black dashed line) is enhanced by 20 times comparing with that of the QDs outside of the MD (red solid line) at RT, and an enhancement of 35 times was observed at 20 K. In general, MDs with a large diameter of around 150 $\mu$m have a high density of WGMs and the mode separations are very small (less than 0.1 nm). Such a small mode separation is less than our spectrometer resolution of about 0.1 nm. Therefore we cannot resolve the different modes of MD as before \cite{Xie2016}. Normally, the cavity modes in different wavelengths could have different quality factors, which results in a different enhancement in the PL spectrum range of the QDs. However, the inset of Figure 3(b) shows the normalized PL spectra of different positions, which are in exact same line shape. This verifies that the cavity enhances the PL emission with the same degree for the emission wavelength range of CIS QDs.

As a comparison, SiO$_{2}$/Si wafers have been used as substrates to drop-cast QDs for optical spectroscopy measurement. Figures 4(a) and 4(b) describe the temporal evolutions of PL spectra of close-packed CIS/ZnS QD solids on SiO$_{2}$/Si substrate with excitation power of about 300 $\mu$W in air and vacuum, respectively. The PL spectra are clearly affected by the laser irradiation time and the environment. In air, the PL intensity initially decreases rapidly and then gradually stabilizes as shown in Figure 4(a). The overall PL intensity reduces to around 10 \% with an irradiation time of 3000 s. It is well known that the decreasing PL intensity of colloidal QDs is due to the photooxidation and the photobleaching \cite{Sark2001,Sykora2010}. To separate photooxidation and photobleaching effects, similar PL measurement was performed in vacuum as shown in Figure 4(b). It can be seen that the PL intensity decreases in a slower manner comparing with that in air. Without considering the photooxidation, the PL decay as a function of time is due to the photobleaching. The photobleaching can be explained by the factor that optically induced damage of oleylamine (OLA) introduces defect states at the surface of CIS/ZnS QDs, resulting in nonradiative channels in the recombination.

To further show the photooxidation induced PL decay behavior, the time-evolutions of normalized PL intensity with different excitation power under air and vacuum are shown in Figures 4(c) and 4(d) respectively. In general, the higher excitation power, the faster decay of the PL intensity both in air and in vacuum. For the sample exposed in air, the PL decay rate increases with increasing the excitation power up to 120 $\mu$W and then saturates, but the decay rate in the vacuum increases monotonically. At a fixed value of excitation power, the PL intensity in air decreases faster than that in vacuum as discussed above. For example, at an excitation power of 300 $\mu$W, the PL intensity drops more than 80 \% in the first 400 s and continues to drop with a slower rate within the rest of the measuring time. While in vacuum, the PL intensity decreases by around 30 \% after 400 s, and no clear saturation can be observed within the measurement time of about 3000 s.

Similarly, time-dependent PL measurement has been performed on the edge of MDs where the PL emission is greatly enhanced, as shown in Figure 5, at different atmospheric environments (air or vacuum) and different temperatures (RT or 20 K). Comparing with Figure 4(c) for QDs exposed in air, similar PL decay is observed at high excitation power. For instance, the PL intensity under excitation power of 150 $\mu$W drops to 80\% after exposure of about 200 s in air and then stabilizes. In vacuum or under low temperature as shown in Figure 5(b) and Figure 5(c), the behavior of PL decay is similar to the case when QDs are on the SiO$_{2}$/Si substrate in vacuum (Figure 4(d)). The higher excitation power the faster decay of PL, which further confirms the photobleaching process. It should be noticed that with low excitation power, the photobleaching effect becomes negligible, for instance, 0.03 $\mu$W in vacuum at RT and 0.37 $\mu$W at low temperature as shown by the top traces in Figures 5 (b) and (c), respectively.

To understand the cavity enhanced emission for CIS/ZnS QDs on MD, we first consider that QDs on the MDs are coupling with the modes of the cavity, which enhances the extraction PL efficiency of QDs \cite{Ganesh2007}. Since the MDs in this work have a large diameter, the cavity modes cover a broad wavelength range including both the emission wavelength range of QDs from 550 to 850 nm and the excitation wavelength at 532 nm. When the excitation laser illuminates with a high excitation power on the edge of MD with a laser spot of 1$\sim$2 $\mu$m in diameter, photoluminescence can be observed from the QDs on the edge of the whole MD, as shown in Figure 2 (a). This confirms that the pumping laser couples to the modes in the cavity and circulating in the cavity which excites all the QDs on the edge of the MD. Additionally, the emitted photons from QDs under the illumining spot coupling with the WGMs of the MD is extracted more efficiently, resulting in an enhanced emission.

Another reason for the enhanced emission is Purcell effect where the QDs couple with the cavity modes in a weakly coupling regime \cite{Gupta2013}. In this case, the spontaneous emission rate is enhanced by suppressing the luminescence lifetime of QDs, which enhances PL emission. Here, no clear increase was observed in slope of the emission intensity versus excitation intensity, which is due to the amplified spontaneous emission. This might be caused by the enhancement of photooxidation, photobleaching and Auger effect with increasing pumping power \cite{Sun2016}. With a reduced lifetime of QDs, the fast Auger process (around 100 ps) could also be suppressed, resulting in an increased biexciton emission in QDs \cite{Giebink2011}. This needs to be confirmed with further measurement using ultrafast time-resolved PL measurement. Nevertheless, excitation power can be greatly reduced to achieve same emission intensity with QDs on the edge of MDs, comparing with that on SiO$_{2}$/Si substrates. The excitation power can be reduced from 7.5 $\mu$W to 0.75 $\mu$W to achieve the luminescence intensity of 1 million counts/s with QDs on the MDs, as shown in Figure 5(d). It can be seen that the rates of the photooxidation and photobleaching are greatly reduced because of the smaller excitation power in the vacuum at RT. One striking observation is that photoinduced degradation almost disappears with a small excitation power while a relative high emission intensity is being obtained with an excitation power of 0.01 $\mu$W on MDs in vacuum at RT. The integrated PL intensity decreased only by 2.1 \% within the measurement time, as shown by inverted blue triangles in Figure 5(d), which still can be improved. But for QDs on SiO$_{2}$/Si substrates, optically induced PL intensity decrease by 6.3 \% is observed for similar PL intensity since it requires a pumping power of 1 $\mu$W (green triangles in Figure 5(d)). No PL degradation in vacuum makes CIS/ZnS QDs very promising for high efficient photon emission for quantum information processing \cite{Michler2000,Brokmann2004,Wissert2011,Xu2004}.

The PL decay of CIS/ZnS QDs can be fitted by a biexponential form as A$_{1}$exp(-t/$\tau$$_{1}$)+A$_{2}$exp(-t/$\tau$$_{2}$) \cite{Wang2003}. Two time constants $\tau_{1}$ and $\tau_{2}$ can be extracted by the fitting PL decay in Figures 4 and 5. Figure 6 summarizes the time constants of the decay of integrated PL intensity as a function of excitation power with a continuous irradiation. The fast decay constant $\tau_{1}$ is in the range of 10 to 160 s, but the $\tau_{2}$ varies in the range of 300 to 3700 s. At RT, both $\tau_{1}$ and $\tau_{2}$ in air are much smaller than those in vacuum, which means the oxygen in the air induces mainly the photooxidation and photobleaching as reported before \cite{Sark2001}. As mentioned before, PL decays faster with increasing the excitation power. Here, the required excitation power on the MDs is much smaller than that on SiO$_{2}$/Si substrates where it has similar decay rates. This is due to that the pumping laser is coupled to the cavity modes, which enhances the excitation efficiency and accelerates the photooxidation process in CIS QDs. It should be noticed by contrast that photoinduced degradation is surprisingly small with a low excitation power, which gives a very long decay time constant $\tau_{2}$. The time constants of PL decay of the sample at 20 K are much smaller than those in vacuum at RT, but are larger than those in air. This could be owing to the fact that the OLA ligands are more easily being damaged by laser at low temperature, resulting in a faster PL decay.

Under a continuous illumination with a pumping laser, temporal evolutions of the peak position and linewidth of PL spectra are very useful to understand the mechanism of photoinduced degradation \cite{Sark2001,Nazzal2004}. The variations of PL peak position and the full width at half maximum (FWHM) as a function of irradiation time are depicted in Figure 7 and Figure 8, respectively. Figure 7(a) represents the PL peak shifting of QDs on the edge of a MD under different excitation power in air. Since the photodegradation here is an irreversible process for QDs, the laser spot was moved to fresh places for each measurement when the excitation power is changed. It can be seen that the start peak position has a blueshift with increasing the excitation power because of the state filling as reported before \cite{Sun2016}. At a low excitation power up to 7.5 $\mu$W, the peak position shows a quick blueshift firstly and then becomes stabilizing as a function of exposure time. As excitation power is further increased, the peak starts with a blueshift quickly (less than 10 s) and then a redshift slowly. In the end, when the excitation power is more than 120 $\mu$W only redshift can be observed within the time resolution in our measurement.

Because of the power induced broadening, the linewidth of PL spectra increases generally with increasing excitation power without considering the time evolutions, as shown in Figure 8(a). In order to see the clear change of the PL spectra, we define $\Delta\lambda$ and $\Delta$FWHM as the difference between the first spectrum and the rest spectra of the central emission peak \cite{Pankiewicz2015}. The $\Delta\lambda$ and $\Delta$FWHM as a function of excitation time with different pumping powers are shown in Figures 7(b)-7(f) and Figures 8(b)-8(f) respectively. The PL blueshifts at a low excitation power exposed in air can be explained by photooxidation. During the illumination exposed in air, the surface of CIS QDs is oxidized when oxygen diffuses through the passivating layer ZnS, which reduces the average sizes and broadens the size distribution of QDs \cite{Manner2012}. The reduced sizes of QDs induce a blueshift because of the increased quantum confinement, which is similar to the results of CdSe/ZnS QDs \cite{Sark2001}. The broadened size distribution induces a linewidth increase of the PL spectra, as shown in Figure 8(b) for the low excitation power cases less than 30 $\mu$W.

One possible reason for redshift could be due to the thermal expansion of QDs because of the increased sample temperature at a high excitation power. However, a surprising phenomenon is that the linewidth decreases with excitation power over 120 $\mu$W as shown in Figures 8(a) and 8(b). The redshift and the linewidth narrowing at high excitation power cannot be simply attributed to thermal expansion. Additionally, the linewidth narrowing with increasing the pumping power could be due to the enhanced coupling between QDs and MDs similar to lasering effect. But this can also be excluded that enhanced spontaneous emission should not be necessarily coming along with a redshift. Therefore, we ascribe the redshift with linewidth narrowing to the agglomeration of QDs induced by high intensity laser illumination. Because of the laser induced OLA ligands damage, the average distances between QDs are reduced after the agglomeration. The reduced average distance enhances the F$\ddot{o}$rster resonant energy transfer (FRET) between different QDs. The enhanced FRET from small QDs to large ones gives a redshift and a narrowed linewidth of PL spectrum. Similar results on both blueshift and redshift can be observed for the QDs on SiO$_{2}$/Si substrates, which are not as strong as the cases on the MDs, as shown in Figure 7(c) and Figure 8(c). Although there are no linewidth narrowing can be observed of the QDs on SiO$_{2}$/Si substrates, the linewidth broadening is slowing down when the excitation power is over 150 $\mu$W (green triangles in Figure 8(c)). This strengthens the claim that MDs accelerate the photooxidation with a high excitation power.

While in vacuum, both blueshift and redshift still can be observed with increasing excitation power on SiO$_{2}$/Si substrates and on MDs, as shown in Figures 7(d) and 7(e). The photooxidation induced blueshift can be ascribed to the absorption of oxygen on the QDs surface in vacuum. The overall peak changes on both blueshift and redshift are much less than those in air. In particular, a redshift is only about 3 nm with QDs on MDs in vacuum, comparing with a 10 nm shift in air. This confirms that oxygen in the air is a key issue in photooxidation and photobleaching. No linewidth narrowing has been observed, as shown in Figures 8(d) and 8(e). It should be emphasized that the PL peak shift and linewidth broadening can be negligible for QDs on MDs in vacuum with a small excitation power, for instance, 0.37 $\mu$W in Figures 7(d) and 8(d). Figure 7(f) shows the PL peak shifts of QDs on the edge of MDs at 20 K as a function of time with different excitation powers, which are much less than those at RT (Figures 7(b) and 7(d)). As the excitation power increases up to 60 $\mu$W, the blue shifts become stronger, but are getting weaker with a further increase of excitation power. This can be explained in a similar way as the case in vacuum. The linewidths do not change very much at 20 K (Figure 8(f)). In contrast to the case in vacuum at RT, the ligands of QDs are more easily being damaged, resulting in faster photobleaching at 20 K. However, no strong redshift can be resolved even with a high excitation power, which shows that the QDs do not aggregate very much at low temperature.

To achieve a high fluorescence intensity, it is necessary to increase the pumping laser power. While due to the photooxidation and photobleaching, the PL intensity of CIS QDs drop-casted on SiO$_{2}$/Si substrates decreases because of the low collection efficiency within the irradiation time both in air and in vacuum. This can be solved by drop-casting QDs on MDs in vacuum, where the photodegradation can be avoided with a low excitation power because of the Purcell effect and the enhanced collection efficiency. Thus, a proper surface protection of CIS QDs on MDs could be used to obtain enhanced PL emission without photodegradation.
\section{Conclusions}
In summary, the temporal evolution of PL properties of CIS/ZnS QDs have been systematically investigated by drop-casting QDs on SiO$_{2}$/Si substrates and MDs with different atmospheric conditions. In air, strong photodegradation has been observed because of the photooxidation and photobleaching accompanying with PL peak blueshift and linewidth broadening at low excitation power. With the further increase of the excitation power, the PL spectra have shown a redshift and narrowing linewidth. This is due to that the photoinduced ligand damage induces the agglomeration of QDs, which enhances FRET between different QDs. The enhanced FRET gives a redshift and a reduced linewidth. All these photobleaching effects are reduced in vacuum, which shows that the oxygen plays an important role in the optically induced PL decay. When the QDs drop-casted on MDs, the pumping laser and emitted photon energy couple with the cavity modes of the MDs. When the pumping laser couples with leaky modes in MDs, the excitation efficiency is greatly enhanced, which accelerates photobleaching. When the QDs couple with cavity modes, however, the extraction efficiency has been greatly improved because of Purcell effect, resulting in a PL enhancement by 20 times at RT and by 35 times at 20 K. By taking advantages of enhanced PL extraction efficiency, the photobleaching can be avoided with a small excitation power but with a relative high PL intensity. Since the photodegradaiton has been one of the major obstacles for the colloidal QDs in many applications, the high efficient PL emission without photodegradation is very promising for utilizing CIS QDs as high efficient photon emitters at RT. This paves a new way for CIS QDs to implement single-photon emission at RT for quantum information processing, instead of InAs QDs working at cryogenic temperature \cite{Xu2004,Xu07}.

\section{Methods}

The CIS/ZnS QDs were synthesized by thermolysis with a mixture solution of Copper(I) acetate, Indium acetate and oleylamine (OLA) as ligands in a high boiling point solvent at 240 $^{\circ}$C. The details of the synthesis process can be found in our previous reports\cite{Chen2012,Bai2016}. The CIS/ZnS core/shell QDs have diameters of about 2$\sim$5 nm and exhibit a tunable PL emission wavelength range from 500 nm to 900 nm \cite{Zhong2012}. The CIS QD powder was dissolved in toluene assisted with ultrasonic vibration, then drop-casted on a SiO$_{2}$/Si substrate or a silica MD \cite{Bo2014}. After the toluene was evaporated, the close-packed QD solids were formed \cite{Pan2015}. Steady-state PL spectra were measured by a conventional confocal micro-PL system, as shown in Figure 1. The CIS QD samples were placed in a cold-finger cryostat with an optical window for PL measurement, with which the sample temperature can be tuned from 20 K to RT. An excitation laser with a wavelength of 532 nm was focused on samples by using a large numerical aperture objective with a spot size around 1$\sim$2 $\mu$m in diameter. The objective is mounted on a two-dimensional PZT controlled stage for spatial mapping. The excitation laser power was measured after the objective above the samples. The fluorescence from CIS/ZnS QDs was collected by the same objective and then dispersed through a 0.55 m spectrometer. The PL spectrum was measured by a charge coupled device camera cooled with liquid nitrogen. The inset of Figure 1 shows a scanning electron microscope (SEM) image of the cross section of a silica MD. The diameter of the MD is around 150 $\mu$m, which is marked in the Figure 1. The WGMs are confined at the edge of the MD which induces high quality factors.The quality factors in the wavelength range of QDs are about 10$^{6}$, by a transmission measurement using a tunable semiconductor laser around 635 nm~\cite{Bo2014}.

A 2D simulation has been performed to calculate the mode distribution for a microdisk with a radius of 75 $\mu$m using finite-difference time-domain (FDTD) method. A refractive index of 1.456 was used, which should be suitable for the emission wavelength range of CIS QDs, from 600 nm to 800 nm. A magnetic dipole (correspond to TE mode) at 563 THz with a pulse width of 3.5 fs was put 74.2  $\mu$m away from the center. Simulation region is a square with a side length of 160 $\mu$m and the mesh is 25 nm in both x and y directions. A mode position at 669.75 nm has been selected to calculate the electromagnetic field distribution, which is close to the peak position of PL emission from QDs, as shown in Figure 2 (d).

\begin{acknowledgements}

This work was supported by the National Basic
Research Program of China under Grant No.
2013CB328706 and 2014CB921003; the National
Natural Science Foundation of China under Grant No. 91436101, 61675228 and 61275060; the
Strategic Priority Research Program of the
Chinese Academy of Sciences under Grant No.
XDB07030200.

\end{acknowledgements}


\newpage
\begin{figure}
  \centering
\includegraphics[scale=0.5]{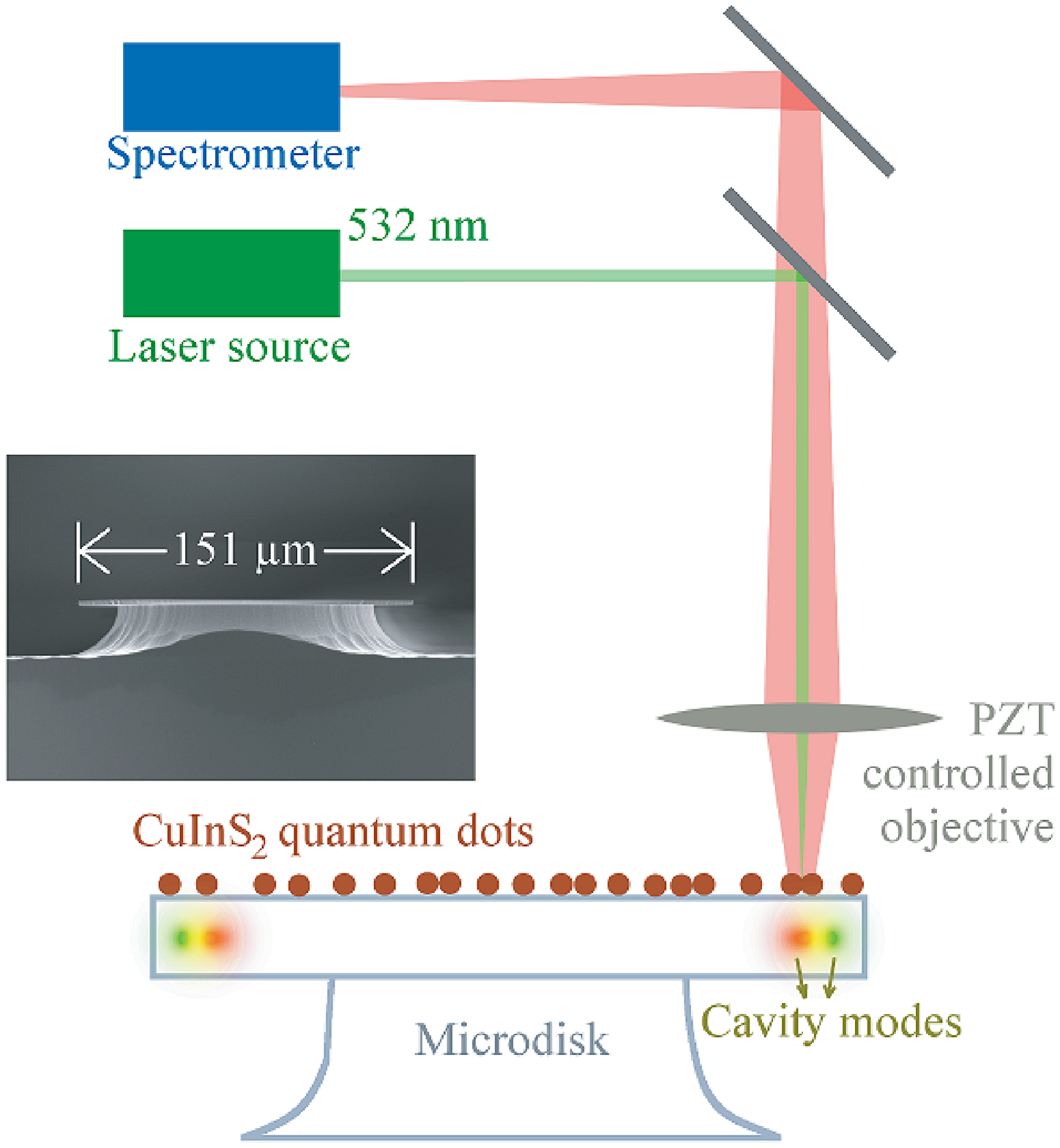}
\caption{Schematic diagram of a confocal micro-PL system for QDs on MDs. The objective lens is mounted on a two-dimensional PZT controlled stage for PL mapping. A scanning electron microscopy image of cross section of a typical MD is shown in the inset.}
\label{Confocal1}
\end{figure}

\newpage
\begin{figure}
  \centering
\includegraphics[scale=0.8]{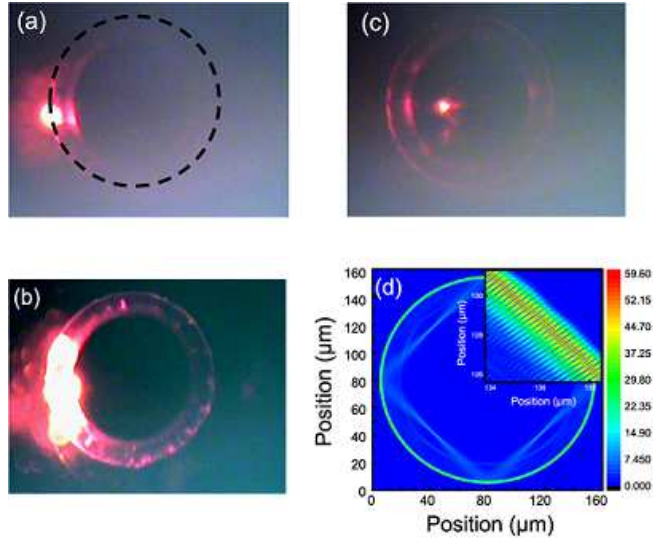}
\caption{Wide-field PL images of CIS QDs on a MD by moving the pumping laser spot crossing the edge of the MD, close to the edge (a), on the edge (b) and away from the edge (c). The pumping laser was filtered when the images were captured. (d) Magnitude of electric field distribution of a cavity mode at 669.75 nm of the MD, which was simulated by a finite-difference time-domain method with following parameters, r = 75 $\mu$m, n$_{SiO_2}$ = 1.456, mesh = 25 nm. The inset shows an enlarged pattern of the mode distribution.
}
\label{Confocal4}
\end{figure}

\newpage
\begin{figure}[htp]
\centering
\includegraphics[scale=0.6]{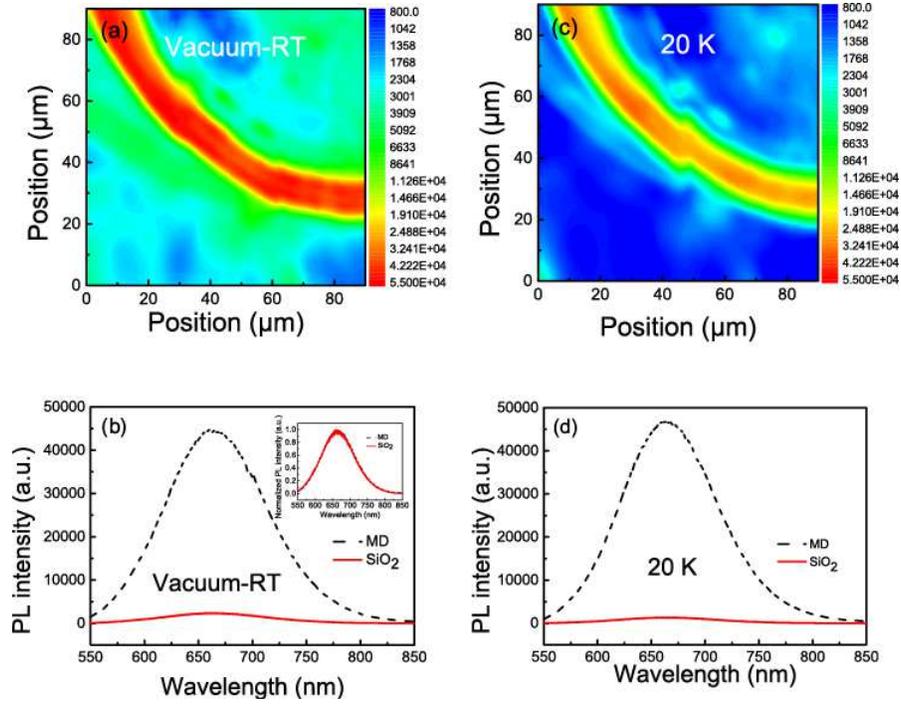}
\caption{(a) and (c) PL mapping of CIS/ZnS QDs on MDs with a scanning area of 80 $\times$ 80 $\mu$m$^2$, the color bars are in log scale, (b) and (d) Typical PL spectra of close-packed CIS QDs solids at RT and at 20 K. The black dashed lines are the PL spectra of QDs on the edge of MD and the red solid lines are PL spectra collected far away from the MD. The inset in (b) shows the normalized PL spectra of QDs. }

\end{figure}

\newpage
\begin{figure}[htp]
\centering
\includegraphics[scale=0.5]{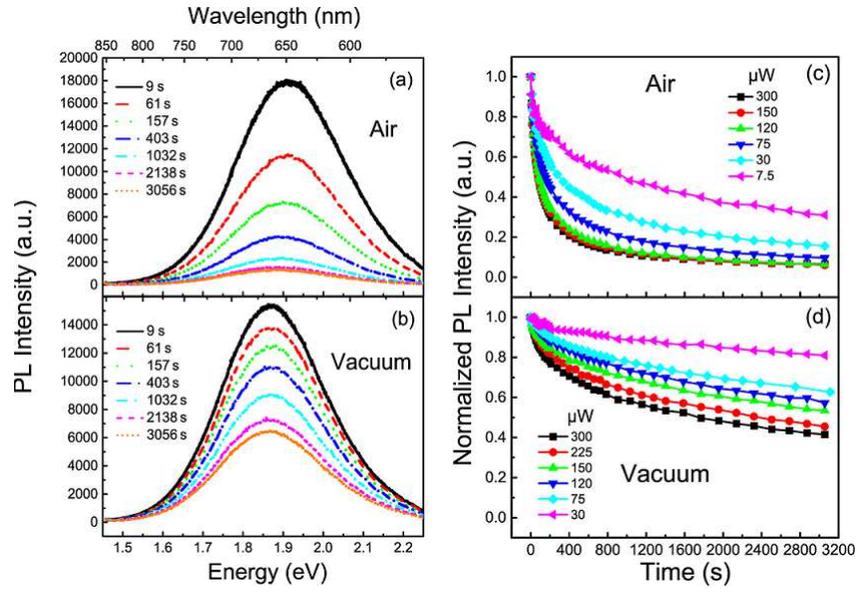}
\caption{PL spectra of the close-packed CIS/ZnS QDs on SiO$_{2}$/Si substrates as a function of irradiation time in air (a) and in vacuum (b). (c) and (d) Normalized PL intensity decay as a function of time for the CIS/ZnS QDs on SiO$_{2}$/Si substrates under different excitation power in air and in vacuum. }
\label{Confocal5}
\end{figure}

\newpage
\begin{figure}[htp]
\centering
\includegraphics[scale=0.5]{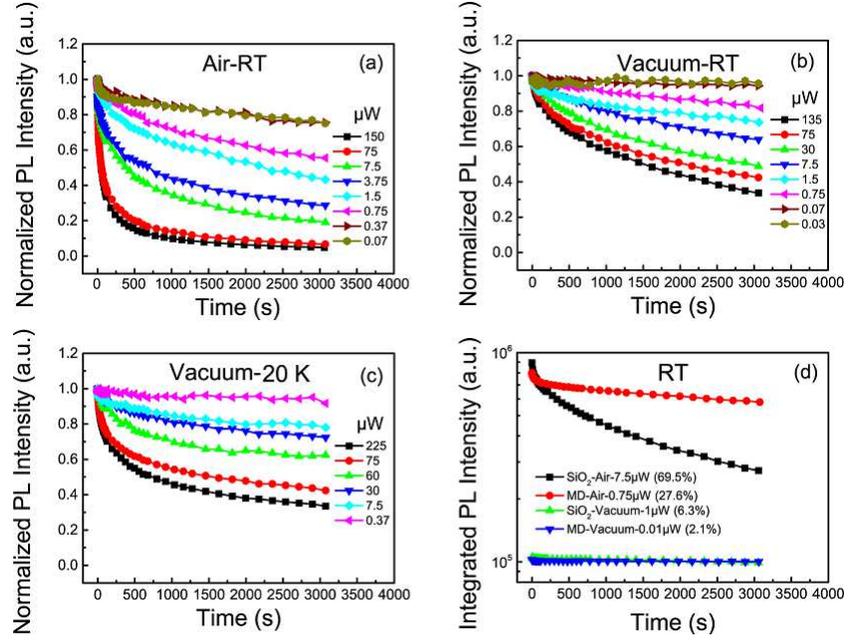}
\caption{Normalized PL decay as a function of time for CIS/ZnS QDs on the edge of MDs under different excitation power in air (a), in vacuum (b) at RT, and at low temperature of 20 K (c). (d) Comparison on the temporal evolutions of integrated PL intensities for QDs on SiO$_{2}$/Si substrates and MDs. The excitation power for each substrate was tuned to achieve similar PL intensities in air and in vacuum. The PL intensity decay percentage for each line is labeled in the figure legend. }
\end{figure}

\newpage
\begin{figure}[htp]
\centering
\includegraphics[scale=0.5]{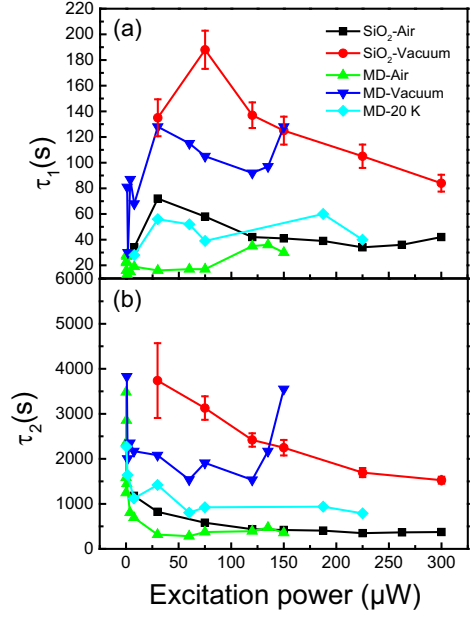}
\caption{Fitted time constants $\tau_{1}$ (a) and $\tau_{2}$ (b) of PL decay as a function of excitation power for QDs on SiO$_{2}$/Si substrates and MDs in air or in vacuum at RT or at 20 K. The error bars are marked only for QDs on a SiO$_{2}$/Si substrate in vacuum to avoid the complexity.}
\label{Confocal7}
\end{figure}

\newpage
\begin{figure}[htp]
\centering
\includegraphics[scale=0.6]{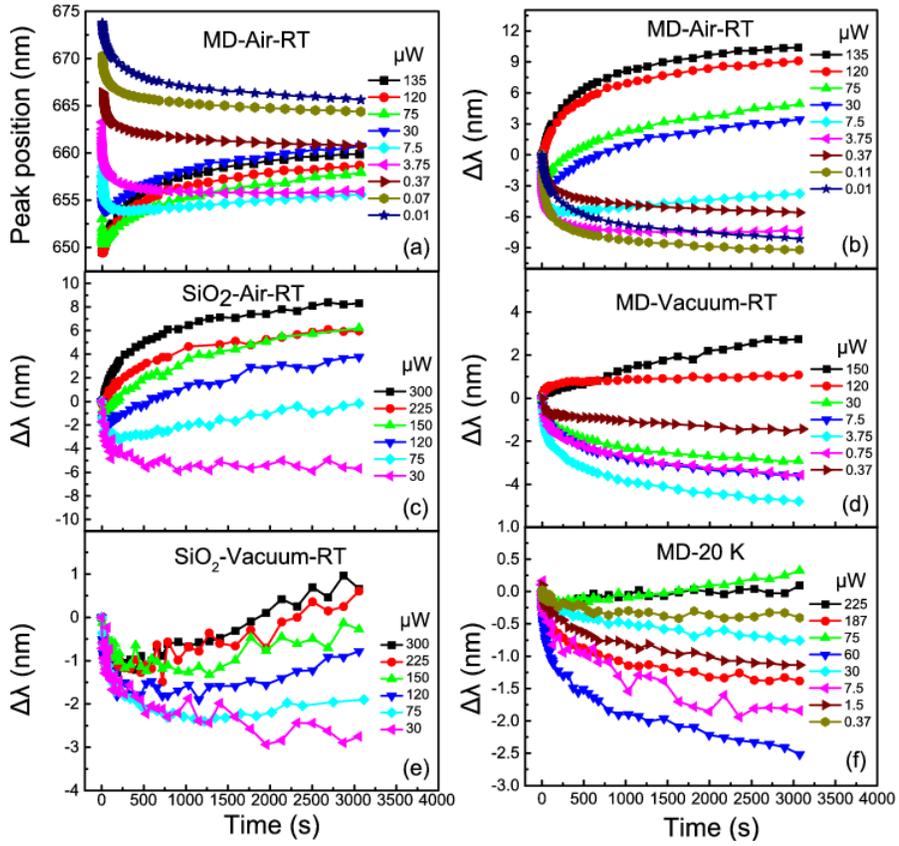}
\caption{(a) PL peak positions as a function of time for QDs on MDs in air at RT. (b)-(f) Temporal evolutions of PL peak position change ($\Delta\lambda$) with various excitation power. The measurement conditions are marked in each panel.}

\end{figure}

\newpage
\begin{figure}[htp]
\centering
\includegraphics[scale=0.6]{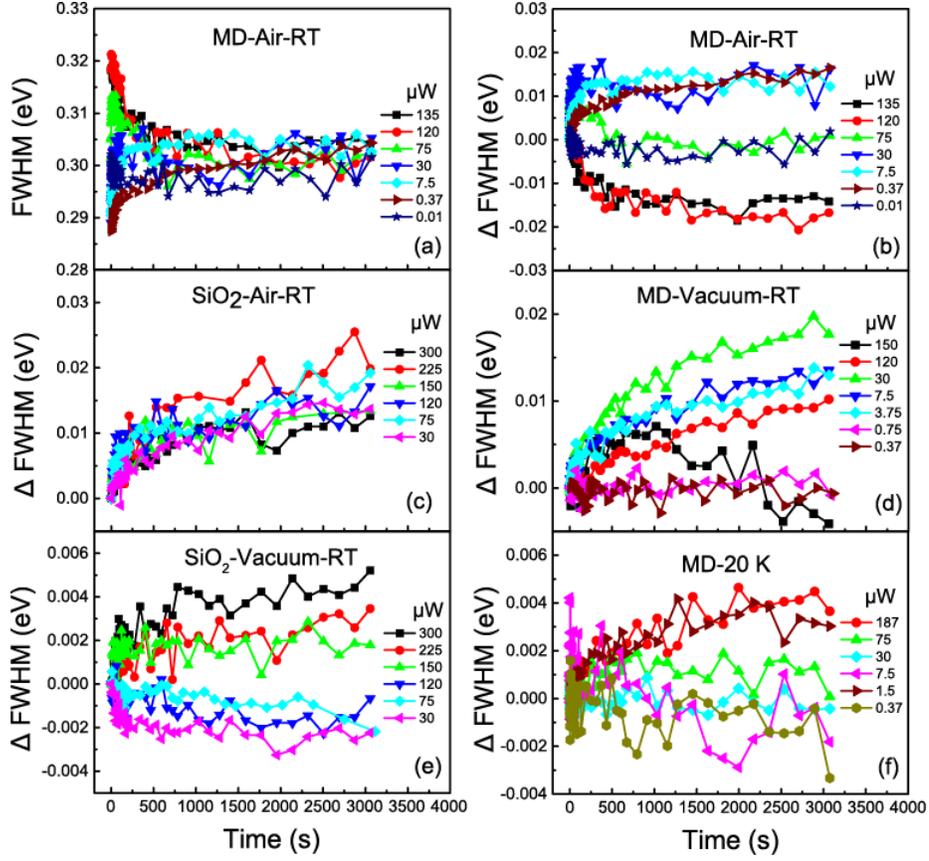}
\caption{(a) PL linewidths as a function of time for QDs on MDs in air at RT. (b)-(f) Temporal evolutions of linewidth change ($\Delta$FWHM) with various excitation power. The measurement conditions are marked in each panel.}

\end{figure}

\end{document}